\documentclass[aps,prb,preprint,floatfix]{revtex4-1}

\usepackage{subfig}

\usepackage{amsmath}  % needed for \tfrac, \bmatrix, etc.
\usepackage{amsfonts} % needed for bold Greek, Fraktur, and blackboard bold
\usepackage{graphicx} % needed for figures
\usepackage{epstopdf} % fixes overprinting of tex with figure?

\begin{document}

\title{The External Magnetic Field Created by the Superposition of
  Identical Parallel Finite Solenoids}

\author{Melody Xuan Lim}
\email{mxl2@duke.edu}
\thanks{Corresponding author}
\author{Henry Greenside}
\email{hsg@phy.duke.edu}
\affiliation{Department of Physics, Duke University, Durham
NC 27708-0305}

\date{\today}

\begin{abstract}
  Using superposition and numerical approximations of an analytical
  expression for the magnetic field generated by a finite
  solenoid, we show that the magnetic field external to parallel
  identical solenoids can be nearly uniform and substantial, even when
  the solenoids have lengths that are large compared to their
  radii. We study two arrangements of solenoids---a ring of parallel
  solenoids whose surfaces are tangent to a common cylindrical surface
  and to nearest neighbours, and a large finite hexagonal array of
  parallel solenoids---and summarize how the magnitude and uniformity
  of the resultant external field depend on the solenoid length and
  distances between solenoids. We also report some novel results about
  single solenoids, e.g., that the energy stored in the internal
  magnetic field exceeds the energy stored in the spatially infinite
  external magnetic field for even short solenoids. These results
  should be broadly interesting to undergraduates learning about
  electricity and magnetism as novel examples of
  superposition based on a familiar source of magnetic fields.
\end{abstract}

\maketitle

\section{Introduction}
\label{sec:intro}

In introductory physics courses on electricity and
magnetism\cite{Young2011} and in upper-level undergraduate courses on
electrodynamics\cite{Griffiths99,Purcell2013}, students learn a
striking fact which is that the magnetic field~${\bf B}$ generated by
a sufficiently long cylindrical solenoid is highly uniform inside and
of small magnitude outside. An implication of this fact is that the
external magnetic field generated by a solenoid can usually be ignored
as small.

But this well-known result raises some questions that are often not
discussed in undergraduate courses. One question is how long must a
solenoid actually be for the external magnetic field to be smaller
than some specified value, at different points in space? And might
there be circumstances in which the external magnetic field of several
solenoids could combine to be become substantial in magnitude, even
for long solenoids? If so, might there also be configurations of
solenoids such that their external magnetic field is also
approximately uniform, so that a single solenoid or a pair of
Helmholtz coils~\cite{Griffiths99} is not the only way to generate an
approximately uniform magnetic field in some region of space?

\begin{figure}[ht]
  \centering
\includegraphics[width=4in]{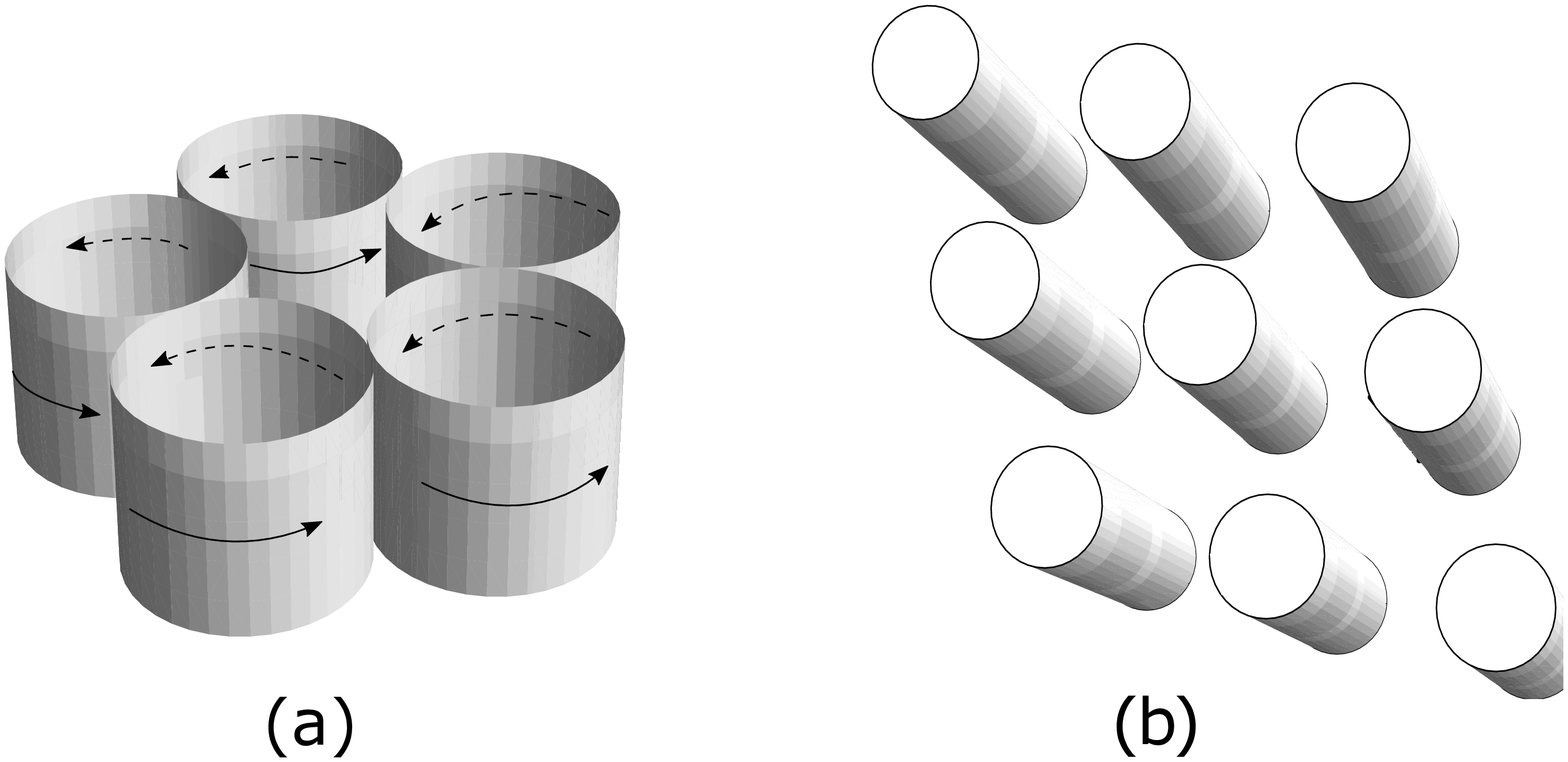}
  \caption{Perspective drawings of two geometric arrangements of parallel identical solenoids. a) A ring of 5 solenoids, each of which is tangent to nearest neighbors and to a common inner cylindrical surface. The direction of the constant current density~$\iota$ on the surface of the solenoids is indicated by the arrows. b) A section of an extended hexagonal array of parallel solenoids. }
\label{fig-geometries}
\end{figure}

In this paper, we investigate these questions for two geometric
arrangements of identical parallel solenoids (see Fig.~\ref{fig-geometries}), a ring of solenoids each
of which is tangent to nearest neighbors and to a common inner
cylindrical surface, and an extended hexagonal array. For both cases,
we find that there is a range of parameters (spacing of the solenoids
and length of the solenoids for a given current density) for which the
external magnetic field is substantial in magnitude, i.e., comparable
to the magnitude of the field at the center of a long
solenoid. Further, for a large hexagonal array of solenoids, the
magnetic field between the solenoids is highly uniform and decreases
slowly in magnitude with increasing solenoid length.

As part of our analysis, we also discuss some calculations and
insights regarding the magnetic field of a single finite solenoid. For
example, we calculate how the external magnetic field of a single
solenoid decreases with increasing solenoid length for points ranging between close to and far from the solenoid's surface and find that a simple analytical
expression derived for the far-field of a solenoid on its symmetry
plane (see Eq.~(\ref{eq-Griffiths-far-field-of-solenoid}) below) gives
a surprisingly accurate approximation of the magnetic field even close
to the surface of the solenoid, at least for solenoids whose lengths
exceed about four radii. This result unifies several previous discussions. We also show that
the energy associated with the internal magnetic field of a solenoid
exceeds the energy associated with the infinitely extended external
magnetic field once the solenoid length exceeds about the radius of
the solenoid, so comparing the internal and external magnetic energies
does not lead to a new length scale that could be used to distinguish
long from short solenoids (see Fig.~\ref{fig-external-frac-energy}
below).

While most of our results are new, parts of our
   Section~\ref{sec:single-solenoid} overlap with some previous
   papers. Brown and Flax\cite{Brown64} discuss a way to calculate the
   magnetic field of a thick solenoid starting from the same integrals
   we use~\cite{Callaghan1960} for a zero-thickness
   solenoid, and Derby and Olbert~\cite{Derby09} has a closely related
   discussion that discusses a simple code to approximate these
   integrals using elliptic functions. Farley and
   Price~\cite{Farley01} show that the magnetic field strength just
   outside and on the midplane of a long finite solenoid of arbitrary
   constant cross-section falls off as $L^{-2}$ with increasing~$L$
   but these authors do not discuss, as we do for solenoids of
   circular cross section, how the magnetic field behaves for short
   solenoids or for points that are at intermediate distances from the
   solenoid's surface. Muniz et al.\cite{Muniz15} used Taylor
   expansions of the magnetic scalar potential to calculate the
   off-axis internal magnetic field of a finite solenoid, which
   complements our direct approximation of two one-dimensional
   integrals using numerical integration, although our method works
   for any point in space.  However, these earlier papers do not
   mention some of our single solenoid results such as how the
   external magnetic energy compares with the internal magnetic
   energy, or how there is a simple expression that provides an
   accurate approximation of the magnetic field on the midplane of a
   solenoid, for all points, not just close to the solenoid's
   surface. We also do not know of previous discussions about the
   magnetic field created by multiple parallel solenoids, which is the
   key contribution of this paper.

Our results should be of broad interest to undergraduate physics
students and to instructors of undergraduate physics courses as a
moderately more complicated example of superposition of magnetic
fields, based on an example---a finite solenoid---that students have
already learned about. The example we discuss below for a ring of
parallel solenoids shows that just having a high symmetry of sources
(many identical solenoids tangent to an inner cylinder) is not enough
to guarantee uniformity of the external magnetic field. The second
example of a large hexagonal array of parallel solenoids also shows
that the external magnetic field of solenoids can be substantial, even
for solenoids whose lengths are large compared to their
radii. Finally, this paper provides a useful example to share with
students of how software like Mathematica~\cite{Mathematica15},
Maple~\cite{Maple15}, and Matlab~\cite{Matlab15} makes it easy for
undergraduates to study numerically and visually superpositions of
magnetic sources that are complicated by their number or geometry.

The rest of this paper is organized as follows. In the next
section, we discuss a previously published analytical result\cite{Callaghan1960, Derby09} for the
magnetic field generated by an idealized
cylindrical continuous solenoid and some insights about this field
based on numerical studies of the analytical solution. In
Section~\ref{sec:ring-of-solenoids}, we use the results of Section~\ref{sec:single-solenoid} to discuss the properties of the
magnetic field generated by a ring of identical parallel continuous
solenoids, while in Section~\ref{sec:hex-array-of-solenoids} we
discuss the properties of the magnetic field generated by a large
hexagonal grid of identical parallel continuous solenoids. After
summarizing key points in Section~\ref{sec:conclusions}, we discuss in
Appendix~\ref{app-validation} how various numerical calculations were
validated, and in Appendix~\ref{app-adding-radial-components} how the
radial components of the magnetic fields from different solenoids were
combined.

\section{The Magnetic Field of a Single Finite Continuous Solenoid}
\label{sec:single-solenoid}

The starting point for our study of the external magnetic field
created by superimposing the fields of several parallel finite
solenoids is an analytical expression\cite{Callaghan1960, Derby09},
detailed below, for the magnetic field~${\bf B}$ generated by a continuous
solenoid, by which which we mean a cylindrical surface of radius~$a$
and length~$L$ such that a spatially uniform time-independent
one-dimensional current density~$\iota$ (with units of amperes per
meter) flows azimuthally around the surface.  

In this section, we discuss some physical properties of this
analytical expression, for example we characterize how the external
magnetic field depends on its length~$L$, and confirm and extend an
analytical result below that the external
magnetic field at some point in space first increases and then
decreases with increasing~$L$, with an asymptotic function behavior
of~$L^{-2}$.  We also compare the energy associated with the external
magnetic field with the energy associated with the internal magnetic
field as a function of~$L/a$ and find that the two energies become
comparable when~$L \approx a$, so that there is not an interesting new
length scale for solenoids related to the magnetic energy.

\begin{figure}[ht]
 {\includegraphics[width=6in]{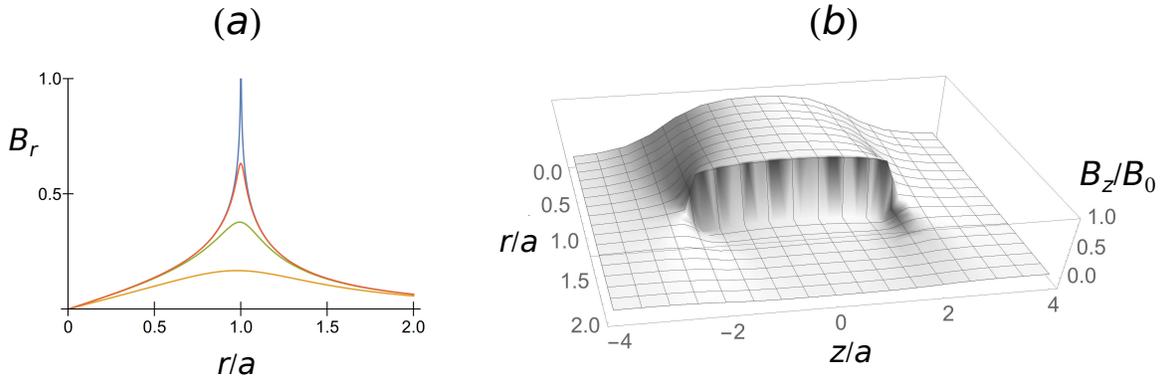}}
\caption{The magnetic field~${\bf B}(r,z)$ of a finite continuous
  solenoid has two discontinuities, here shown for parameters~$a=1$, length~$L=4$, and unit current
  density~$\iota=1$. (a) The radial component~$B_r(r,z)$ diverges
  for~$z=\pm L/2$ at~$r=a$. Here we show four curves of~$B_z(r,z)$
  for~$z=0.8(L/2), 0.95(L/2), 0.99(L/2)$, and~$L/2$ (from lowest to highest
  curves). The curve for $z=L/2$ (and similar curve for $z=-L/2$) shows a
      divergence of~$B_r$ at $r=a$. One can show from Eq.~(2) that
      this is a logarithmic divergence of the form $\ln(|r-a|)$ for
      $r$ sufficiently close to~$a$ and for~$z=\pm L/2$.
  (b)~The axial
  component~$B_z(r,z)$ has a jump discontinuity for~$-L/2 \le z \le
  L/2$ at~$r=a$ that represents the change in direction and magnitude
  of the magnetic field as one progresses radially from just inside to just
  outside the surface of the solenoid on a line of constant~$z$.}
\label{fig-B-field-discontinuities}
\end{figure}

A continuous solenoid is an idealization of real solenoids that are
helically wound with wires of some finite thickness. The continuous
solenoid has the advantage over real solenoids of requiring just three
parameters to specify---a radius~$a$, length~$L$, and uniform current
density~$\iota$---and has the further advantage that the analytical
expression for its magnetic field is easy to evaluate numerically,
accurately and quickly, at any point in space. The continuous solenoid
is somewhat unphysical in that the zero-thickness of the surface
causes the radial component of the magnetic field to diverge in
magnitude at the top and bottom edges of the solenoid's surface (see
Fig.~\ref{fig-B-field-discontinuities}(a)), while no such divergence
occurs for physical solenoids of finite thickness. However, we show
below in this section that this divergence is not important since it
still leads to a finite amount of energy stored in the magnetic field,
and that the region where~$B_r$ has a large magnitude has a small
spatial extent (compared to the solenoid's radius) so that the
magnetic field of the continuous solenoid provides quantitatively
useful insights about the magnetic field of physical solenoids.

To describe the axisymmetric magnetic field of a continuous solenoid
of radius~$a$, length~$L$, and current density~$\iota$, we introduce a
cylindrical coordinate system~$(r,\theta,z)$ such that~$r=0$ defines
the solenoid's axis and the plane~$z=0$ bisects the solenoid (so the
ends of the solenoid lie at the coordinates $z=\pm L/2$). For this coordinate system, the current density~$\iota$ flows counter-clockwise, so that the solenoid's internal magnetic field~${\bf B}$ points in the positive~$\hat{\bf z}$ direction. From the azimuthal symmetry of the problem, we deduce that~$B_{\theta}=0$, and furthermore that~$B_z$ and ~$B_r$ do not depend on~$\theta$, so that the magnetic field has the form 

\begin{equation}
  \label{eq-B-in-cyl-coordinates}
    {\bf B} = B_r(r,z) \hat{\bf r} + B_z(r,z) \hat{\bf z} .
\end{equation}

Calculations
then show~\cite{Callaghan1960, Derby09} that the magnetic field at a point~$(r,z)$ in space has components given in terms of the
following one-dimensional integrals:
\begin{eqnarray}
  B_r &=& -\frac{a \mu_0 \iota}{2 \pi}
     \int_0^\pi d\alpha \left[
       \frac{\cos\alpha}{
         \sqrt{ \xi^2 + r^2 + a^2 - 2 a r \cos\alpha }}
     \right]_{\xi_-}^{\xi_+} ,   \label{eq-br-nasa-1d-integral} \\
  B_z &=& \frac{a \mu_0 \iota}{2 \pi}
     \int_0^\pi d\alpha \left[
         \frac{\xi \, (a - r \cos\alpha)}{
         \left( r^2 + a^2 - 2 a r \cos\alpha \right)
         \sqrt{ \xi^2 + r^2 + a^2 - 2 a r \cos\alpha } }
     \right]_{\xi_-}^{\xi_+} .   \label{eq-bz-nasa-1d-integral}
\end{eqnarray}
The lengths~$\xi_\pm$ are defined by
\begin{equation}
  \label{eq-xi-pm-defn}
  \xi_\pm = z \pm \frac{L}{2} ,
\end{equation}
and the notation $\bigl[f(\xi)\bigr]^{\xi_+}_{\xi_-}$ is an
abbreviated way of writing $f(\xi_+)-f(\xi_-)$.

Appendix~\ref{app-validation} gives details of how we approximated
Eqs.~(\ref{eq-br-nasa-1d-integral}) and~(\ref{eq-bz-nasa-1d-integral})
numerically for given values of~$r$, $z$, $a$, $L$, and~$\iota$ using
Mathematica~\cite{Mathematica15}. This appendix also summarizes how we
validated the numerical approximations of these integrals, and results
obtained for superpositions of solenoids.
% by comparing equivalent expressions that involve elliptic
%integrals and verifying that various analytically know limits for the
%magnetic field could be accurately reproduced. 
We found that Mathematica's numerical approximations to
Eqs.~(\ref{eq-br-nasa-1d-integral}) and~(\ref{eq-bz-nasa-1d-integral})
had a relative accuracy that exceeded eight significant digits, except
for the component~$B_r$ near the coordinates~$(r,z)=(a,\pm L/2)$,
where this component diverges to infinity. This accuracy was more than
adequate for our calculations.

\begin{figure}[ht]
  \centering
  \includegraphics[width=4in]{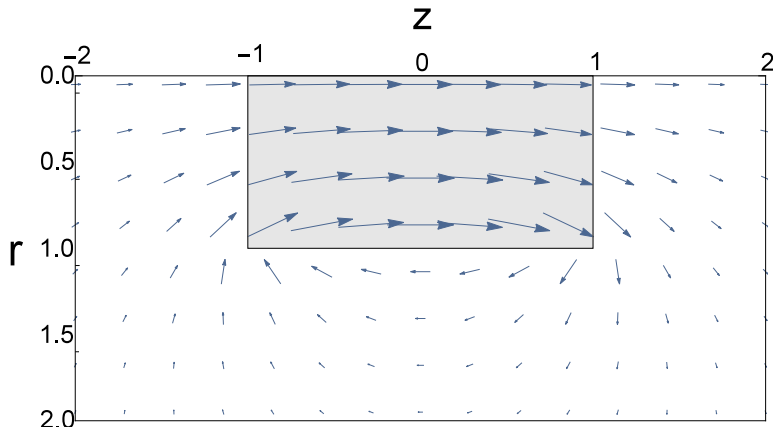}
  \caption{Vector plot of the axisymmetric magnetic field~${\bf
      B}(r,z)$ generated by a finite continuous solenoid,
    Eqs.~(\ref{eq-B-in-cyl-coordinates})-(\ref{eq-bz-nasa-1d-integral}),
    for radius~$a=1$, length~$L=2$, and current
    density~$\iota=1$. Vectors are shown only for~$r \ge 0$ so only
    the lower half of the solenoid (gray region) and half of the magnetic field
    are shown.  The internal field is approximately uniform despite
    the solenoid's short length. Even for this short length, the interior magnetic field has attained a magnitude~$B \approx 0.8B_0$ at the solenoid's center~$r=z=0$, where~$B_0$ is defined in Eq.~(\ref{eq-B-infinity-defin}). }
\label{fig-vector-B-field-single-solenoid}
\end{figure}

Fig.~\ref{fig-vector-B-field-single-solenoid} shows a vector-field
plot of the magnetic field~${\bf B}$ generated by
Eqs.~(\ref{eq-br-nasa-1d-integral}) and~(\ref{eq-bz-nasa-1d-integral})
for a solenoid of radius~$a=1$, length~$L=2$, and current
density~$\iota=1$.  Even for this short length, the interior magnetic
field is approximately uniform and the magnetic field at the
solenoid's center~$r=z=0$ has attained a magnitude~$B \approx 0.8B_0$
that nearly equals the magnitude
\begin{equation}
  \label{eq-B-infinity-defin}
  B_0 = \mu_0 \iota ,
\end{equation}
of the uniform internal magnetic field of an infinitely long solenoid
with the same current density~$\iota$. We note that
Eq.~(\ref{eq-B-infinity-defin}) is the familiar
formula\cite{Young2011} $B=\mu_0 n I$ for the magnetic field magnitude
inside a long solenoid that is uniformly wound with~$n$ wires per unit
length and with each wire carrying a current~$I$. The product $nI$ has
units of current per length and corresponds to the current
density~$\iota$ in the limit $n \to\infty$, $I \to 0$ with~$nI$
finite.

% The external magnetic field approximates that of an equivalent point
% magnetic dipole~$\boldsymbol{\mu} = m \hat{\bf z}$ located at the
% center~$r=z=0$ of the solenoid, with moment $m = (\pi a^2) \iota L$.
% (This value for~$m$ can be deduced by comparing
% Eq.~(\ref{eq-Griffiths-far-field-of-solenoid}) below with the
% magnetic field of a point dipole $\mu_0 m / (4 \pi r^3)$ in the
% far-field limit~$r \gg L, a$ for~$z=0$.)  
%
% More details: see Eq. (5.86) on page 246 of Griffiths 3e which gives
% the far-field magnetic field of a point dipole for theta=pi/2, m/(4
% pi r^3). Comparing this with the expression given in Griffiths page
% 254 for exercise 5.61 for the far field of a rotating charged
% cylindrical surface, in the limit s >> a, L. The current density
% iota is related to the charge density sigma and angular frequency
% omega by iota = a omega sigma (current I in small vertical segment
% of length dz is [(2 pi a)dz]sigma / T so iota = omega a sigma.

It is difficult to see from
Fig.~\ref{fig-vector-B-field-single-solenoid} that the magnetic field
of a finite continuous solenoid has two kinds of
discontinuities. Fig.~\ref{fig-B-field-discontinuities}(a) shows how
the radial component~$B_r(r,z)$ varies with~$r$ for fixed~$z$, for
several different values of~$z$ that approach a singularity at~$r=a$
that occurs for~$z=\pm L/2$. (So~$B_r$ diverges everywhere on the two
ends of the solenoid at~$z=\pm L/2$, which is again an artifact of our
approximating the solenoid as a zero-thickness surface; this
singularity does not occur for a solenoid of finite thickness with a
corresponding volumetric current density~$J$.) By setting $z=L/2$ in Eq. (2), one can evaluate the integral
       analytically over the range $0 \le r \le a-\epsilon$ and then
       find that the integral has a term proportional to
       $\ln(\epsilon)$ and so diverges logarithmically in the limit $\epsilon
       \to 0$. This will imply that the total energy associated with
       the magnetic field (Eq. (7) below) is a finite quantity.

The surface plot of the axial component~$B_z(r,z)$ in
Fig.~\ref{fig-B-field-discontinuities}(b) shows a more familiar
discontinuity, that the $z$-component~$B_z$ of the magnetic field
abruptly changes direction and magnitude as one passes radially from just
inside to just outside the surface of the solenoid. (This is also
apparent from Fig.~\ref{fig-vector-B-field-single-solenoid} for~$r$
near~$r=1$ and~$z=0$, you can see how the magnetic field~${\bf B}$
changes direction and magnitude near the surface of the solenoid.)

We next discuss how the external magnetic field of the finite
continuous solenoid decreases with increasing length~$L$ for fixed
radius~$a$ and for fixed current density~$\iota$, an important issue
for our discussion in later sections regarding how large can the
magnitude of the external magnetic field be for multiple parallel
continuous solenoids of some given length~$L$. Here there is some
prior analytical insight via Exercise~5.61 on page~254 of Griffith's
textbook on electrodynamics~\cite{Griffiths99}, which states that the
magnetic field on the solenoid's midplane ($z=0$) far from the axis
($r \gg a$) asymptotically has the form ${\bf B} = - B \hat{\bf z}$,
where the magnitude~$B$ is
\begin{equation}
  \label{eq-Griffiths-far-field-of-solenoid}
  B \approx 
   \mu_0 \iota \times \frac{L a^2}{4 \big( r^2 + (L/2)^2 \big)^{3/2}} .
\end{equation}
This is the leading-order term in an expansion in powers of the small
quantity $a/r$ and so its validity when~$a/r$ is not small, say close
to the surface of the solenoid, when~$r \simeq a$, is not known.

%A complementary analytical result for the magnetic field close to the
%surface of the solenoid and close to the bisecting plane ($|z/a| \ll
%1$) is provided by Problem~5.5(b) on page~226 of Jackson's text on
%electrodynamics~\cite{Jackson99}, which gives the following
%expression:
%\begin{equation}
%  \label{eq-Jackson-Bz-just-outside-solenoid}
%  B_z(r=a^+,z) \approx
%  - \left(
%      2 \mu_0 \iota a^2 \over L
%    \right)
%    \left(
%    1 + 12 \left(z\over L\right)^2 - 9 \left(a \over L\right)^2 +
%    \cdots 
%    \right) ,
%\end{equation}
%where the dots~$\cdots$ denote higher-order terms in the small
%quantities ~$|z/L|$ and~$|a/L|$. The validity of this expression is
%not known for modest lengths~$L$, for~$z$ not near zero in magnitude,
%or for points not close to the surface of the solenoid.

Equation~(\ref{eq-Griffiths-far-field-of-solenoid}) makes two
predictions. First, for sufficiently large solenoid lengths~$L \gg r$,
$B_z \propto L^{-2}$ so~$B_z$ decays rather slowly (algebraically
rather than exponentially) to zero with increasing~$L$. We note that
this~$1/L^2$ scaling is also how the magnetic field at the center of
the solenoid, $r=z=0$, converges to its infinite-length value
Eq.~(\ref{eq-B-infinity-defin}) i.e.~$(B_z - B_0)/B_0 \propto L^{-2}$
for sufficiently large~$L$. (This follows from the analytical
expression for the magnetic field on the axis of a finite solenoid,
see Eq.~(\ref{eq-Bz-on-axis-of-solenoid}) in
Appendix~\ref{app-validation}.) Second, if
this expression has validity for general values of~$r$,
Eq.~(\ref{eq-Griffiths-far-field-of-solenoid}) predicts that at a
fixed radial distance~$r$ from the axis, $B$~does not decrease
monotonically with increasing~$L$, but first increases with a local
maximum at~$L_{\rm max} = \sqrt{2} r$,
and then asymptotically decays to zero as~$1/L^2$. 

%Equation~(\ref{eq-Jackson-Bz-just-outside-solenoid}) in turn predicts
%that, near the plane~$z=0$ and sufficiently close to the external
%surface of the solenoid, $B \propto 1/L$, a slower decay in the limit
%of large~$L$. But Melody could not get reasonable agreement when she
%checked the expression with the NASA field

\begin{figure}[ht]
  \includegraphics[width=6.6in]{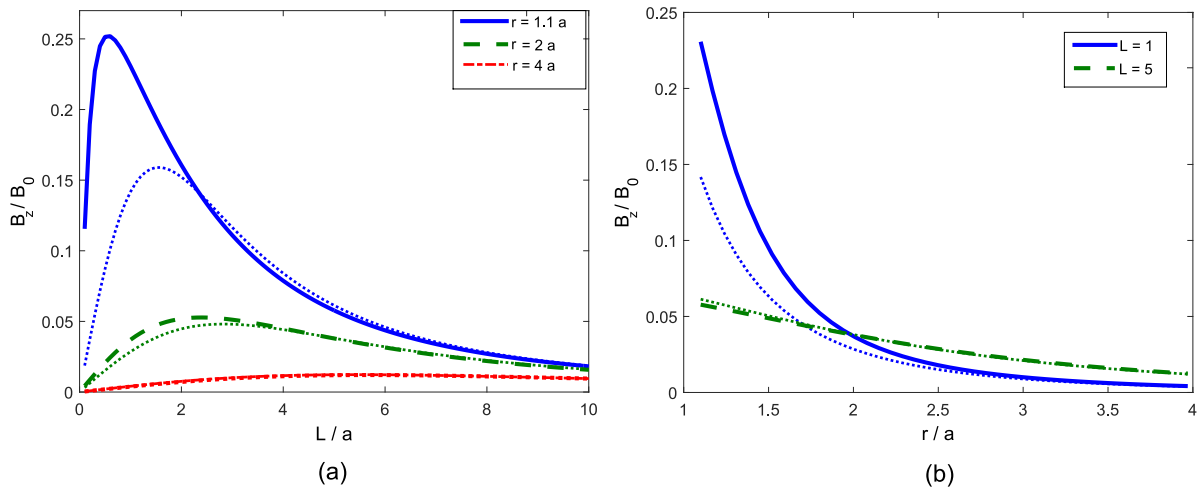}
  \caption{(a) Comparison of Griffiths' result, Eq.~(\ref{eq-Griffiths-far-field-of-solenoid}), with the external magnetic field $B_z(r,z=0)$ of a single finite continuous solenoid, normalized to the
    infinite solenoid axial value~$B_0$ (Eq.~(\ref{eq-B-infinity-defin}))
    versus the solenoid length~$L/a$. Results are shown for three representative radial
    values, with the dotted lines showing Griffiths' corresponding result. Note how~$B_z$ first increases and then decreases with
    increasing~$L$. For~$r=4a$, the analytical result is the same as
    Eq.~(\ref{eq-Griffiths-far-field-of-solenoid}) to the accuracy of
    the plot. (b) The asymptotic result
    Eq.~(\ref{eq-Griffiths-far-field-of-solenoid}) for~$B_z(r,0)$
    derived for~$r \gg a$ (dotted lines) accurately describes the
    radial dependence of the analytical field~$B_z(r)$ even for~$r$
    close to the surface of the solenoid, provided that $L \gtrsim 4a$
    The analytical result and numerical results are indistinguishable
    at the level of these plots for $L/a \ge 7$.  }
  \label{fig-field-outside-single-solenoid-compare-griffiths}
\end{figure}

In Fig.~\ref{fig-field-outside-single-solenoid-compare-griffiths}, we
compare these predictions with the exact analytical field of
Eqs~(\ref{eq-br-nasa-1d-integral}) and~(\ref{eq-bz-nasa-1d-integral})
for radial distances that are close to and far from the solenoid's
surface. Panel~(a) shows that, for the representative fixed radial
values~$r=1.1a$, $2a$ and~$4a$ on the~$z=0$ symmetry plane, the
$z$-component of the magnetic field~$B_z(r,L)$ indeed first increases
and then decreases with increasing~$L$ as suggested by
Eq.~(\ref{eq-Griffiths-far-field-of-solenoid}), even for relatively
short solenoids for which $L/a \gtrsim 2$. Surprisingly, the
$L$-dependence of the analytical field at some fixed~$r$ is accurately
described by Eq.~(\ref{eq-Griffiths-far-field-of-solenoid}) even for
distances close to the solenoid ($r \gtrsim 2a$). Panel~(b) shows a
complementary result, that even for modest fixed solenoid lengths ($L
\gtrsim 4a$), Eq.~(\ref{eq-Griffiths-far-field-of-solenoid}) also
describes the radial dependence~$B_z(r,0)$ well, even for radii close
to the outer surface of the solenoid.  We conclude that
Eq.~(\ref{eq-Griffiths-far-field-of-solenoid}) can be used as a quick
and easy way to determine the magnitude of the magnetic field on the
symmetry plane~$z=0$ of configurations of parallel solenoids without
having to evaluate the integral Eq.~(\ref{eq-bz-nasa-1d-integral}).

\begin{figure}[ht]
\centering
\includegraphics[width=4in]{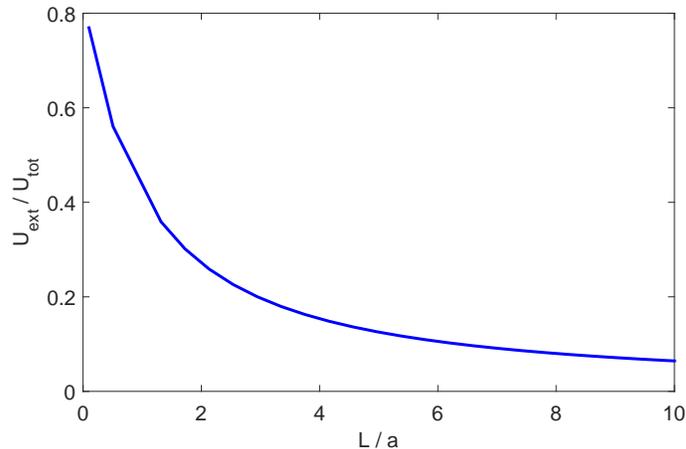}
\caption{Ratio of the energy~$U_{\rm ext}$,
  Eq.~(\ref{eq-magnetic-energy}), associated with the magnetic field
  external to the solenoid ($r > a$ or $|z| > L/2$) to the total
  magnetic energy~$U_{\rm tot}$ over all of space for a single
  continuous solenoid of fixed radius~$a=1$ and fixed current density~$\iota$, as a function of solenoid
  length~$L$. The energy associated with the magnetic field inside the
  solenoid already exceeds the energy associated with the external
  magnetic field for short solenoids such that $L > 0.69 a$.}
\label{fig-external-frac-energy}
\end{figure}

The analytical result Eq.~(\ref{eq-Bz-on-axis-of-solenoid}) for~$B_z$
on the axis of a finite continuous solenoid shows that solenoids whose lengths~$L
\gtrsim 4a$ are already ``sufficiently long'' in the sense that the
magnetic field magnitude~$|B_z(r=0,z=0)|$ at the solenoid's center
already exceeds 90\% of its infinite-length value~$\mu_0 \iota$. It
occurred to the authors that an alternative way to identify when a
solenoid is ``sufficiently long'' would be to ask for what
length~$L^\ast$ (for a given radius~$a$ and given current
density~$\iota$) does the energy~$U_{\rm ext}$ associated with the
magnetic field
\begin{equation}
  \label{eq-magnetic-energy}
  U = \int \frac{B^2}{2 \mu_0} \, d^3{\bf r}
    = \frac{1}{2 \mu_0}
        \int\!\!\int 2 \pi r \, \left( B_r^2 + B_z^2 \right) \, dr \, dz
\end{equation}
external to the solenoid (the infinite region defined by~$r > a$ or
$|z| > L/2$), become less than half of the total energy~$U_{\rm tot}$
associated with the entire magnetic field
(Eq.~(\ref{eq-magnetic-energy}) evaluated over all of space, $r \ge
0$).  By approximating the integral Eq.~(\ref{eq-magnetic-energy}) numerically (see Appendix~\ref{app-validation}), we calculated and show in Fig. \ref{fig-external-frac-energy} the ratio~$U_{\rm
  ext} / U_{\rm tot}$, as a function of solenoid length~$L$ for
parameters~$a=1$ and~$\iota=1$. This ratio falls below the value~$1/2$ for~$L
\approx 0.69 a$ so that, even for quite short solenoids, the energy
associated with the magnetic field inside the solenoid exceeds the
energy associated with the external field, even though the latter has
infinite spatial extent. 

\section{The Magnetic Field of a Ring of Identical Parallel Finite Continuous
  Solenoids}
\label{sec:ring-of-solenoids}

\begin{figure}
  \includegraphics[width=0.4\textwidth]{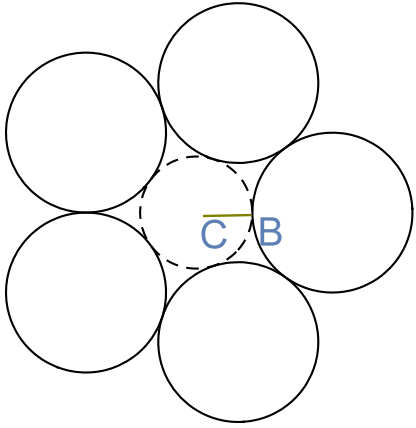}
    \caption{Cross-section of a ring
      of $n=5$ parallel continuous solenoids of length~$L=2$ and
      radius~$r \approx 1.4$ that are tangent to neighboring
      solenoids and to a common inner cylindrical surface of
      radius~$R=1$, shown as a dotted ring with center~$C$.  The line
      segment~CB denotes the location of the rectangular area,
      transverse to the plane of the figure, in which the magnetic
      field~${\bf B}$ is plotted in
      Fig.~\ref{fig-ring-of-solenoids-B-field}. For a perspective drawing, see Fig.~\ref{fig-geometries}(a). }
\label{fig-ring-of-solenoids-geometry}
\end{figure}

We are interested in the question of whether the magnetic field
external to multiple parallel finite solenoids can ever be substantial
in magnitude (comparable to the value~$B_0=\mu_0 \iota$ on the axis of
an infinite solenoid) and approximately uniform. In this
section, we consider an arrangement of~$n \ge 3$ parallel identical finite
continuous solenoids, each with current density~$\iota=1$ and of
length~$L$, that are arranged in a ring as shown in
Figs.~\ref{fig-ring-of-solenoids-geometry} and~\ref{fig-geometries}(a). Each solenoid is tangent to
an inner common cylindrical surface of radius~$R$ and tangent to its
neighbors. Because a substantial portion (nearly~50\% in the case of
many solenoids) of the external magnetic field of each solenoid passes
through the common central cylindrical region, this ring geometry is a
plausible candidate for producing an external magnetic field of
substantial magnitude. We show in this section that this ring geometry
can indeed produce a substantial magnetic field for solenoids that are long compared to their radii, but that the magnetic
field in the common cylindrical region is generally not uniform.  In
the following section, Section~\ref{sec:hex-array-of-solenoids}, we
discuss similar questions for a large finite hexagonal array of
parallel finite continuous solenoids.

To describe the external magnetic field in the central cylindrical
region, we introduce a second cylindrical coordinate system~$(\rho,\phi,z)$ such
that~$\rho=0$ corresponds to the axis of the inner cylindrical
surface, $\rho=R$ corresponds to the cylindrical surface that all the
solenoids are tangent to, and~$z=0$ is the common bisecting plane of
the parallel solenoids.  For $n$~parallel solenoids that are all
tangent to the inner cylindrical surface of radius~$R$ and tangent to
neighbors, the tangency conditions imply that the common solenoid
radius~$a_n$ is given by
\begin{equation}
  \label{eq-ring-solenoid-radii}
    a_n = \frac{\sin\left(\pi / n \right)}{1 -\sin\left( \pi / n
      \right) } \, R , \qquad n \ge 3 ,
\end{equation}
so $a_3 \approx 6.5R$, $a_4 \approx 2.4R$, $a_5 \approx 1.4R$, $a_6 =
R$, and~$a \approx \pi R/n$ becomes small for large~$n$.
% The geometric argument is that, in cross-section, one has two
% circles of radius~$a$ tangent to each other and tangent to an inner
% circle of radius~$R$. Each solenoid spans an angle~$\theta$ from the
% center of the inner cylinder, so $n \theta = 2 \pi$ or
% $\theta=2\pi/n$. But the tangency conditions give $sin(\theta/2) =
% a/(a+R)$ which gives Eq.~(\ref{eq-ring-solenoid-radii}).

At any point~$(\rho,\phi,z)$ inside the central cylindrical
region~$\rho \le R$, the $z$-component of the magnetic
field~$B_z(\rho,\phi,z)$ is obtained by directly adding the
values~$B_z(r_i,z)$ from each solenoid in the ring at that point,
where~$r_i$ is the distance of the point~$(\rho,\phi,z)$ to the axis
of the~$i$th solenoid. Calculating the radial
component~$B_\rho(\rho,\phi,z)$ is more involved since the radial
direction~$\hat{\bf r}_i$ with respect to the $i$th~solenoid
at~$(\rho,\phi,z)$ varies with~$i$. The details are given in
Appendix~\ref{app-adding-radial-components}.

%\begin{figure}
%  \includegraphics[width=6in]{fig-bz-different-z.eps}
%  \caption{(a) Local unit vectors~$\hat{\bf B}(\rho,z)$ of the
%    external magnetic field produced by a ring of~$n=5$ parallel
%    continuous solenoids of length~$L=2$ that are tangent to an inner
%    cylindrical surface of radius~$R=1$. The vectors lie in the plane
%    defined by~$\phi=0$, corresponding to line segment CB in
%    Fig.~\ref{fig-ring-of-solenoids-geometry}. The magnetic field is
%    not uniform except in a small central region, and becomes even
%    more non-uniform for a ring with more solenoids (larger~$n$).  (b)
%    Perspective drawing of three of the solenoids in a ring of five
%    solenoids. The five dotted segments indicate the different values
%    of~$z$ for which the component~$B_z(\rho,z)$ was plotted as a
%    function of~$\rho$. (c) }
%  \label{fig-ring-of-solenoids-B-field}
%\end{figure}

\begin{figure}
   \includegraphics[width=6.5in]{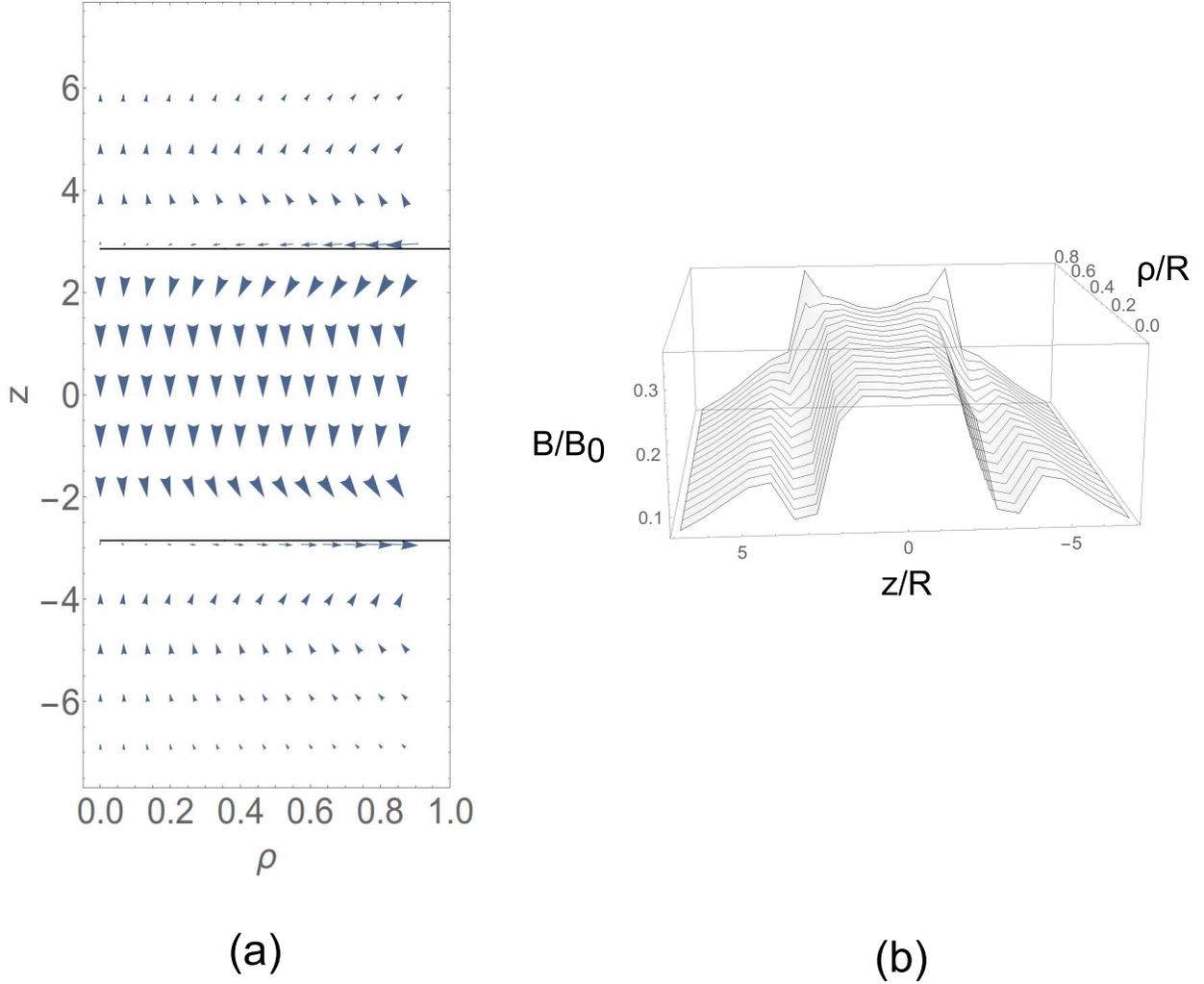}
   \caption{External magnetic field produced by 5~parallel continuous
     solenoids tangent to an inner cylinder of radius~$R=1$, for
     solenoid length~$L=4a \approx 5.6$ where the solenoid radius~$a
     \approx 1.4$ . (a) Vector plot of ${\bf B}(\rho,z)$ in the
     plane~$\phi=0$ (line~CB of
     Fig.~\ref{fig-ring-of-solenoids-geometry}(b)) shows the
     direction and magnitude of the magnetic field inside the central
     cylindrical region. The lengths of the vector heads are
     proportional to the local magnitude~$B$. (b) Surface plot of
     field magnitude~$B/B_0$ normalized to the magnetic field
     magnitude~$B_0$ inside an infinite single solenoid of the same
     current density~$\iota$. }
     \label{fig-ring-of-solenoids-B-field}
\end{figure}

We now discuss our calculations of the magnetic field in the central
cylindrical region for rings with varying number~$n$ of continuous
solenoids and varying length~$L$, for a central cylindrical region of
fixed radius~$R=1$ and for all solenoids having the same current
density~$\iota=1$. 

For a ring of $n=5$~solenoids of radius~$a\approx 1.4$ (see
Eq.~(\ref{eq-ring-solenoid-radii})) and length~$L=4a$,
Fig.~\ref{fig-ring-of-solenoids-B-field}(a) shows a vector plot of the
magnetic field~${\bf B}(\rho,\phi,z)$ inside the central cylindrical
region $0 \le \rho \le R$ along the plane $\phi=0$, which corresponds
to a rectangle that is perpendicular to and passes through the line
segment~CB in Fig~\ref{fig-ring-of-solenoids-geometry}(b). We see that
the magnetic field is approximately uniform in direction and magnitude
with a magnitude~$B/B_0 \approx 0.3$ that is about one third the
magnitude of the magnetic field~$B_0= \mu_0 \iota$ found on the axis
of an infinitely long single solenoid with the same current
density~$\iota$. The magnetic field deviates substantially from that
of a single solenoid (see
Fig.~\ref{fig-vector-B-field-single-solenoid}) in the planes~$z = \pm
L/2 \approx \pm 2.8$, which are indicated by the horizontal black
lines in Fig.~\ref{fig-ring-of-solenoids-B-field}(a); the magnetic
field is purely horizontal and of greatly decreased magnitude. 

Further insight about the structure of the central magnetic field is
provided by panel (b) of
Fig.~\ref{fig-ring-of-solenoids-B-field}, which shows a surface plot of
the field magnitude~$B(\rho,z)$ for the same rectangle as
panel~(a). Panel~(b) confirms that the magnetic field is approximately
uniform inside the central cylindrical region (the magnitude~$B$ is
approximately constant for $|z| \le L/2$) but becomes non-uniform near
the boundary of one of the solenoids because of the divergence
of~$B_r$ near the ends of the solenoid (see
Fig.~\ref{fig-B-field-discontinuities}(a)). 

\begin{figure}
  \includegraphics[width=0.6\textwidth]{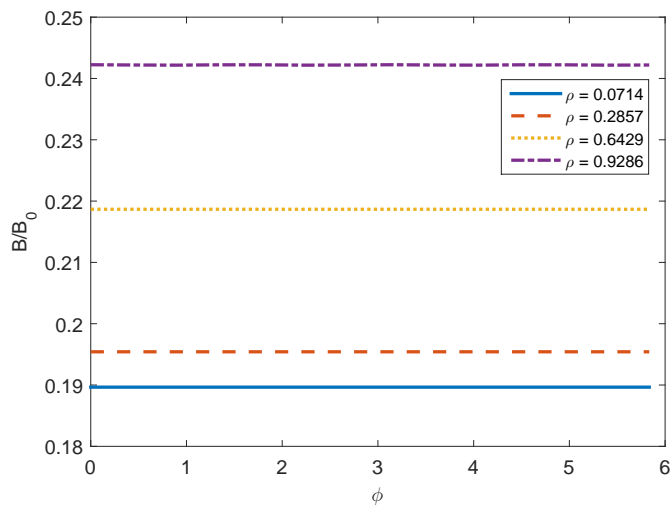}
  \caption{Plot of $B_z(\rho,\phi,z=0)$ inside the inner
    cylindrical surface of an arrangement of five solenoids each
    with~$L/a=4$ and~$\iota=1$, for~$\rho = 0.07, 0.29, 0.64$ and~$0.93$.  The magnetic field has high azimuthal
    symmetry, despite there being five discrete sources of magnetic field. There is only evidence of a five-fold symmetry close to the edges of the solenoids.}
\label{fig-ring-r-theta-plane}
\end{figure}

Fig.~\ref{fig-ring-r-theta-plane} shows that the magnetic field inside
the cylindrical region is highly axisymmetric for~$z=0$. Despite the fact that
the field is being generated by five discrete solenoids, there is graphical evidence of a five-fold symmetry only for~$r$ close to~$R$ amd for~$z \simeq \pm L/2$, as shown in Fig~\ref{fig-ring-of-solenoids-B-field}(b).  

\begin{figure}
  \includegraphics[width=4in]{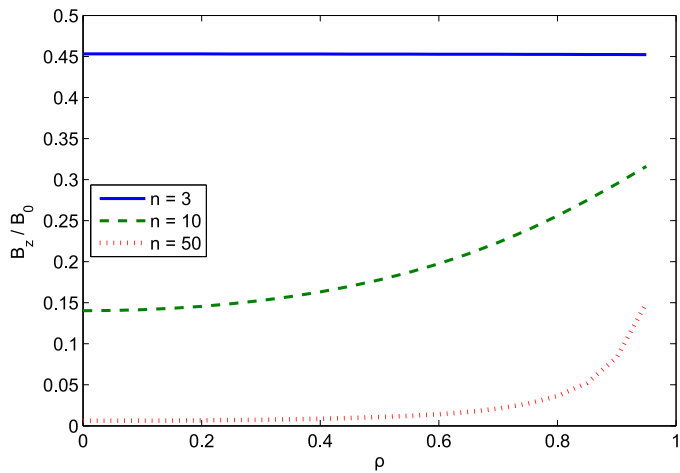}
  \caption{Axial component~$B_z(\rho,z=0)$ on the bisection
    plane~$z=0$ for $n=3$, $10$, and~$50$ solenoids, for fixed
    solenoid length~$L/a=4$ and fixed current density~$\iota$. The
    increase in the number~$n$ of sources cannot compensate for the
    $1/L^2$ decrease in magnetic field strength so the magnetic field
    in the central cylindrical region becomes small for large~$n$.}
  \label{fig-bz-radial-nrings}
\end{figure}

We found that the external magnetic field in the central cylindrical
region becomes more nonuniform and decays radially in magnitude for
larger numbers of solenoids ($n > 4$) so
Fig.~\ref{fig-ring-of-solenoids-B-field} is close to the best case in
terms of achieving a uniform magnetic field in a ring geometry. As the
number~$n$ of solenoids increases for fixed~$R$ and fixed~$L$, the
solenoid radii~$a_n$ Eq.~(\ref{eq-ring-solenoid-radii}) decrease in
size, the solenoids effectively become longer compared to their
radius, and so the magnetic field external to each solenoid decreases
in magnitude according to
Eq.~(\ref{eq-Griffiths-far-field-of-solenoid}), being largest near the
surfaces of the solenoids ($\rho \approx 1$) and decreasing towards
the center of the common cylindrical region where~$\rho=0$.  This is
illustrated in Fig.~\ref{fig-bz-radial-nrings} which shows how
$B_z/B_0(\rho,\phi=0,z=0)$ varies on the bisecting plane~$z=0$ for increasing
values of solenoid number~$n$, for a fixed length~$L/a=4$. 

%Figure~\ref{fig-bz-different-z} shows more quantitatively how the
%radial dependence of~$B_z(r;z)$ depends on the value of~$z$ for a ring
%of 5~solenoids. We find that the uniformity in~$B_z(r,0)$ is localized
%to the plane of symmetry of the solenoids, as well as to several other
%locations, for instance at~$z=0$ and~$z=2L$ a highly uniform magnetic
%field is generated. Furthermore, the change in curvature of the
%functional form of~$B_z{r}$ between $z=L/2$ and $z=3L/4$ suggests that
%there must also exist a region of uniform magnetic field just outside
%of the edge of the ring of solenoids.  Taken together with
%Fig.~\ref{fig-ring-of-solenoids-B-field}, these results suggest
%that~$B_z$ inside the ring of solenoids does not reverse in direction
%until~$z>3L/4$, unlike the axial dependence of~$B_z$ in the single
%solenoid.

\section{The Magnetic Field of an Hexagonal Array of Identical Parallel Continuous Solenoids}
\label{sec:hex-array-of-solenoids}
The second configuration of parallel finite solenoids that we explore
is a large finite hexagonal lattice. Since we have shown in Fig~\ref{fig-field-outside-single-solenoid-compare-griffiths} of Sec.~\ref{sec:single-solenoid} that the external magnetic field of a single solenoid can be as large as~$0.25 B_0$ on axis, just outside the solenoid, this geometry is a potential candidate for producing an external magnetic field of substantial magnitude (at least for specific solenoid lengths and spacings). 
 
We define an
hexagonal Bravais lattice\cite{Ashcroft76} such that the
locations~${\bf v}$ of each solenoid axis in the~$z=0$ plane are given by the
two-dimensional vectors
\begin{equation}
  \label{eq-bravais-lattice}
  {\bf v} =  s \left( m {\bf v}_1 + n {\bf v}_2 \right) + {\bf v}_3 ,
\end{equation}
where the coefficients~$m$ and~$n$ go over all possible integers, and
where the positive parameter~$s$ denotes the lattice spacing (distance
between two nearest neighbor solenoid axes). The basis vectors~${\bf
  v}_1$ and~${\bf v}_2$ are given by
\begin{equation}
  \label{eq-basis-vectors-v1-v2}
  {\bf v}_1 = \hat{\bf i}, \qquad \mbox{and} \qquad
  {\bf v}_2 = \left(\frac{ -1 }{ 2 }\right) \hat{\bf i} 
                  + \frac{ \sqrt{3} }{ 2 } \hat{\bf j}  .
\end{equation}
The vector~${\bf v}_3$, which here is ~$\left( 1/2 \right) \hat{\bf i}  + (1/ 2\sqrt{3}) \hat{\bf j}$, shifts the lattice relative to the origin of
the coordinate system and places the origin at the centroid of the triangle formed by three nearest-neighbour solenoids.  For a given
lattice spacing~$s$, the solenoid radius~$a$ must satisfy $a < s/2$ so 
that that the solenoid surfaces do not intersect.

The array of solenoids is characterized by
the lattice spacing $s$ and the length-to-radius ratio~$L/a$ of each of the
solenoids. We introduce an xyz-Cartesian coordinate system, 
where the origin $(x,y,z)=(0,0,0)$ corresponds to the centroid of the triangle formed by the centers of three nearest-neighbour solenoids in a cell
of the hexagonal lattice. Again, $z=0$ corresponds to the bisecting plane
of the solenoids.

\begin{figure}
\centering
\includegraphics[width=0.7\textwidth]{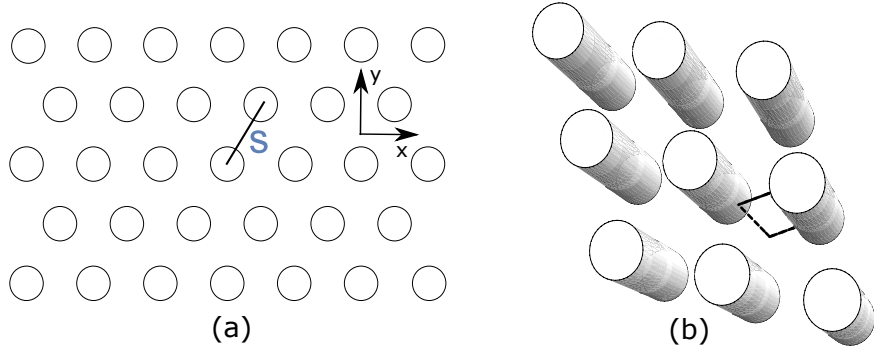}
\caption{ (a) Horizontal cross-section through a portion of an 
  hexagonal lattice of parallel continuous solenoids for lattice
  spacing~$s=1$ as defined by Eq~\ref{eq-bravais-lattice} and~\ref{eq-basis-vectors-v1-v2}. The
  common radius of the solenoids was chosen to be~$a=1/4$. Also shown are the directions of the coordinates~$x$ and~$y$.(b) Perspective drawing of a section of the lattice. The black dashed rectangle corresponds to the
  region in which we compute the magnetic field shown in
  Figure~\ref{fig-hex-lattice-B-vector-field}.}
\label{fig-hex-lattice-geometry}
\end{figure}

\begin{figure}
  \centering
  \includegraphics[width=0.6\textwidth]{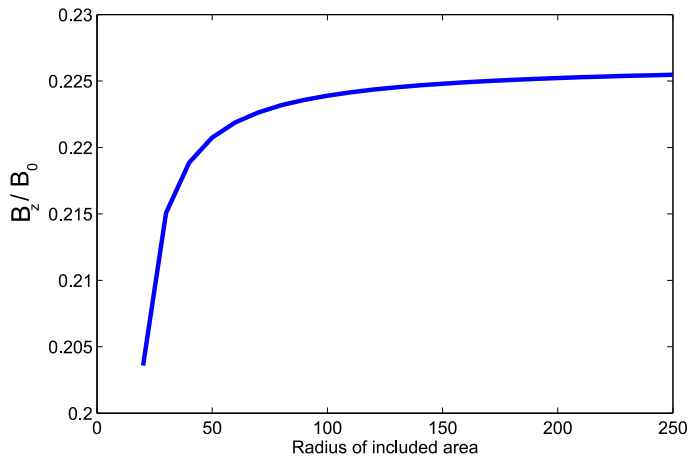}
  \caption{Plot of the magnetic field component $B_z/B_0$ at the centroid
  of three solenoids in an hexagonal lattice as a function of the
  radius~$R$ of a circle centered at the centroid. For the largest radius
  ~$R/s=500$ (not shown), there are $\sim 750,000$ solenoids enclosed within the circle. Note
  how over the range $30<R<500$, $B_z$ varies only be about 10\%, and furthermore appears to asymptote to a finite value.  A
  circle of radius~$=100s$ already approximates the infinite lattice
  case to better than 5\%.}
\label{fig-hex-lattice-B-sum}
\end{figure}

We approximate the total
magnetic field near the origin for an infinite period lattice was found by summing the fields from many solenoids in a
large, finite, approximately circular region centered on~$(0,0,0)$. Fig.~\ref{fig-hex-lattice-B-sum} shows the component~$B_z$ at $(0,0,0)$ as a function of the radius of
the area from which fields of individual solenoids were
summed. Summing the fields from an area of the hexagonal array with
radius~$100s$ is sufficient to approximate the magnetic field due
to the infinite array with a relative accuracy better than~$0.05$.

\begin{figure}
\centering
\includegraphics[width=3in]{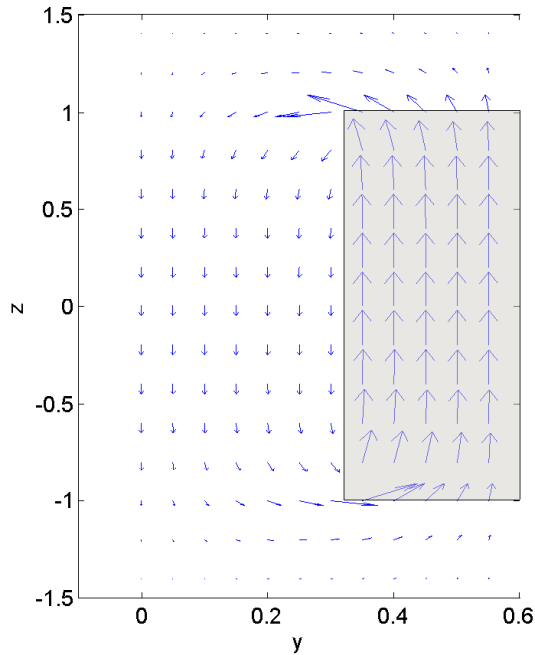}
\caption{Plot of the magnetic vector field~$\bf{B}$ generated between solenoids in a
  hexagonal lattice in the black dashed rectangle shown in
  Fig.~\ref{fig-hex-lattice-geometry}(b), with $s=1$, $a=1/4$, $\iota=1$,
  and~$L/a=8$, summed over solenoids in a circular area with radius~$R=100$. The location of
  the solenoid is indicated by a gray rectangle.  Only the in-plane
  components of the magnetic field are non-zero. The magnetic field is
  highly uniform in the region external to solenoids, for~$|z|<L/2$.}
\label{fig-hex-lattice-B-vector-field}
\end{figure}

The magnetic field at a given point~$(x,y,z)$ was computed by
individually summing the components of the magnetic field from the
solenoids in the lattice from a region with radius~$100s$ (see Appendix~\ref{app-adding-radial-components}). The field within each unit cell of the hexagonal lattice is again axisymmetric. 
Further details of the structure of the magnetic field within a single cell of the lattice
are shown in Fig.~\ref{fig-hex-lattice-B-vector-field}, which is a
vector field plot of the magnetic field in the plane defined by the
line section marked BC in Fig.~\ref{fig-hex-lattice-geometry} and
with axial coordinates between~$z=\pm L$. The external magnetic field away from the edges of the solenoid is comparable in
uniformity to the magnetic field inside the solenoid, with a magnitude~$B_z = 0.6 B_0$. This external magnetic field is quite uniform for most of~$z \leq |L/2|$ , unlike the central magnetic field described in Sec.~\ref{sec:ring-of-solenoids}. 

\begin{figure}
  \centering
  \includegraphics[width=4in]{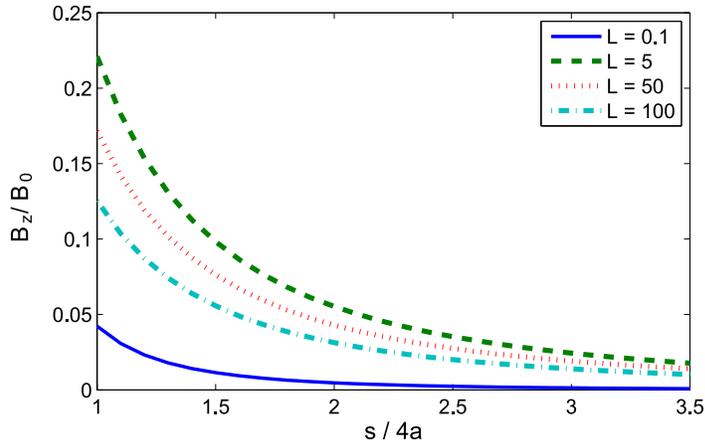}
  \caption{Plot of the ratio $B_z(0,0,0)/B_{0}$ as a function of the
    lattice spacing~$s$ for an hexagonal array of parallel identical
    solenoids of length~$L=0.1,5,50$ and~$100$, radius~$a=1/4$, and current density 
    ~$\iota=1$. The decay of~$B_z$ towards 0 follows
    a power law with a numerically derived exponent that approaches~$-2$ for increasing~$L$, consistent with Eq.~(\ref{eq-Griffiths-far-field-of-solenoid}).}
\label{fig-hex-Bz-lattice-spacing}
\end{figure}

We also found that the magnetic field~$B_z(0,0,0)$ decreases approximately algebraically as a function of the lattice size~$s$ for fixed~$a$ and~$L$. Fig~\ref{fig-hex-Bz-lattice-spacing} shows how the magnetic
field~$B_z(0,0,0)$ decreases in magnitude with increasing lattice
size~$s$ for parameter values $\iota=1$, $a=1/4$, and~$L=0.1,5,50$, and~$100$. The decay of~$B_z$ to 0 follows an approximate power law with a numerical exponent
that approaches~$-2$ with increasing solenoid length. Furthermore, we
find that the magnitude of the magnetic field for any given lattice
spacing depends nonmonotonically on the length of the solenoid. However, this dependence is small: for fixed~$s$, we observe in Fig.~\ref{fig-hex-Bz-lattice-spacing} that changing the lattice spacing by a factor of 20 decreases~$B_z/B_0$ only by a factor of~$\approx 0.5$. 

\begin{figure}
  \centering
  \includegraphics[width=4in]{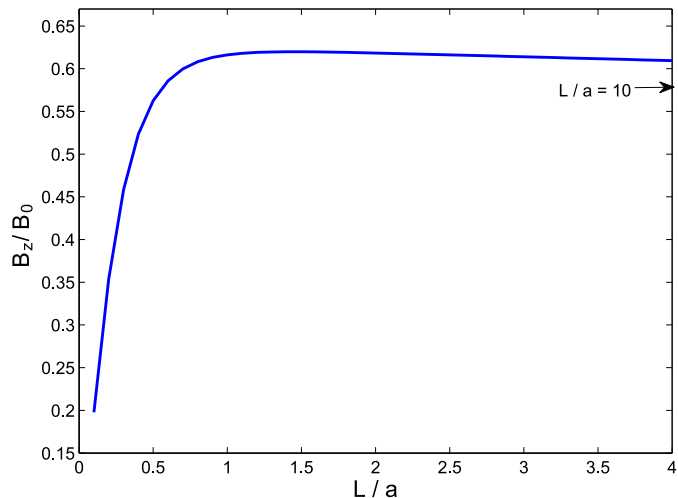}
  \caption{Plot of the ratio~$B_z(0,0,0)/B_0$ as a function of the
    solenoid length~$L$ for an hexagonal array of solenoids. For~$L/a = 10$,~$B_z/B_0$ reaches a value of 0.58, as marked by an arrow. The
    magnetic field attains a local maximum
    at~$~L=2$ of about half the field of a single finite solenoid, after which it decays slowly towards the infinite
    length limit of~$B_z=0$.  }
  \label{fig-hex-Bz-length}
\end{figure}

This slow decay of~$B_z/B_0$ towards zero as a function of~$L$, for fixed~$s$, can be
seen more clearly in Fig.~\ref{fig-hex-Bz-length}, which
shows~$B_z(0,0,0)/B_0$ as a function of $L/a$ for $a=1/4$, $s=1$, and
$\iota = 1$.  As expected from our discussion of the external magnetic
field of the single solenoid, the decay of $B_z$ is
nonmonotonic. However, after reaching a local maximum, $B_z$ decays
linearly, with a numerically derived functional form~$B_z = -0.0045 L
+ 0.63$, from which we extrapolate that~$B_z$ decays to 0 only for a
solenoid length of~$\sim$~140. This linear decay of the external magnetic field of the single solenoid provides another example of a feature of the magnetic field that would have been difficult to predict without carrying out a numerical study. 

\section{Conclusions}
\label{sec:conclusions}

In this paper, we have investigated using elementary methods that
sophomore physics and engineering students can understand (numerical
approximations to one-dimensional integrals and superposition by
direct summation) whether the magnetic field external to some
arrangement of identical parallel finite solenoids of radius~$a$,
length~$L$, and uniform current density~$\iota$ can ever be
substantial in magnitude and possibly uniform. Our motivation was to
provide a deeper insight to the properties of the magnetic field
external to a finite solenoid, which in many undergraduate physics
textbook is ignored because its magnitude is known to become
negligibly small for solenoids that are sufficiently long compared to
their radii (although what is meant by ``sufficiently long'' is rarely
discussed).

We were also interested in the question of whether the supposedly
small external magnetic field could ever be made substantial (compared
to the magnetic field magnitude Eq.~(\ref{eq-B-infinity-defin}) inside
an infinitely long solenoid of identical current density) and possibly
highly uniform by combining the magnetic fields of many parallel
identical finite solenoids. This question is not easy to answer, say
by applying Amp\`ere's law to some small rectangular loop as is
commonly done in undergraduate physics textbooks to a single infinite
solenoid\cite{Young2011,Griffiths99}, since one does not know in
advance the symmetry of the magnetic field, e.g., whether it is
everywhere parallel to the axes of the solenoids, and uniform in
magnitude.

We studied two configurations of identical parallel solenoids with
high symmetry since high symmetry should favor uniformity of the
external field. One configuration consisted of a ring of~$n$ parallel
solenoids that each are tangent to nearest neighbors and to a common
inner cylindrical surface (see
Fig.~\ref{fig-ring-of-solenoids-geometry}). The second configuration
was a large, finite, roughly circular hexagonal array of parallel
solenoids (see Fig.~\ref{fig-hex-lattice-geometry}), that we used to
approximate the magnetic field generated by an infinite hexagonal
array of parallel solenoids. For both cases, we investigated how the
external magnetic field depended on the radius, length, and spacing of
the solenoids, while the current density was kept fixed.

For the ring geometry and hexagonal array, we found that the external
magnetic field could have a substantial magnitude, even for solenoids
that are long compared to their radii. But only for the magnetic field
generated by a large hexagonal array did we find that the external
magnetic field could also be highly uniform, at least in regions of
space that lies between the two planes that contain the ends of the
solenoids.

The starting point for our analysis was an exact analytical
expression,
Eqs.~(\ref{eq-B-in-cyl-coordinates})-Eq.~(\ref{eq-bz-nasa-1d-integral}),
in terms of two one-dimensional integrals for the axisymmetric
magnetic field~${\bf B}$ produced by a single finite continuous
solenoid. It turns out that these one-dimensional integrals can be
rapidly and accurately approximated at all points in space (except for
points near the edges~$r=a$ and~$z=\pm L/2$ of the solenoid, where the
radial component~$B_r$ diverges logarithmically) to better than eight
significant digits using simple and straightforward invocations of
functions provided by the mathematics program
Mathematica\cite{Mathematica15} (see
Fig.~\ref{fig-mma-NIntegrate-code-for-bz} as an example). The same
Mathematica program provided easily used tools to add up the magnetic
fields of many solenoids by superposition, and then visualize the
resulting magnetic field. 

Thus this paper, in addition to providing new insights about the
magnetic field generated by a single solenoid and by groups of
solenoids, should provide a useful example to share with undergraduate
physics students about how mathematics software environments like
Mathematica can be used to explore physics problems that are
substantially more challenging and more interesting than what can be
analyzed by analytical methods. Based on our experience, we feel that
such software should be more widely and systematically incorporated
into undergraduate physics courses so that students learn how to use
analytical, numerical, and visual ways to understand or explore a
variety of physics problems.

\appendix 

\section{Validation of the numerical calculations}
\label{app-validation}

The Mathematica code needed to carry out the calculations discussed
in previous sections is modest, about four pages in length. But
it took some effort to determine whether the results were correct and their accuracy. 
We summarize in this section some of the steps we took to validate the
results.

\begin{figure}[ht!]
  \centering
  \includegraphics[width=6.5in]{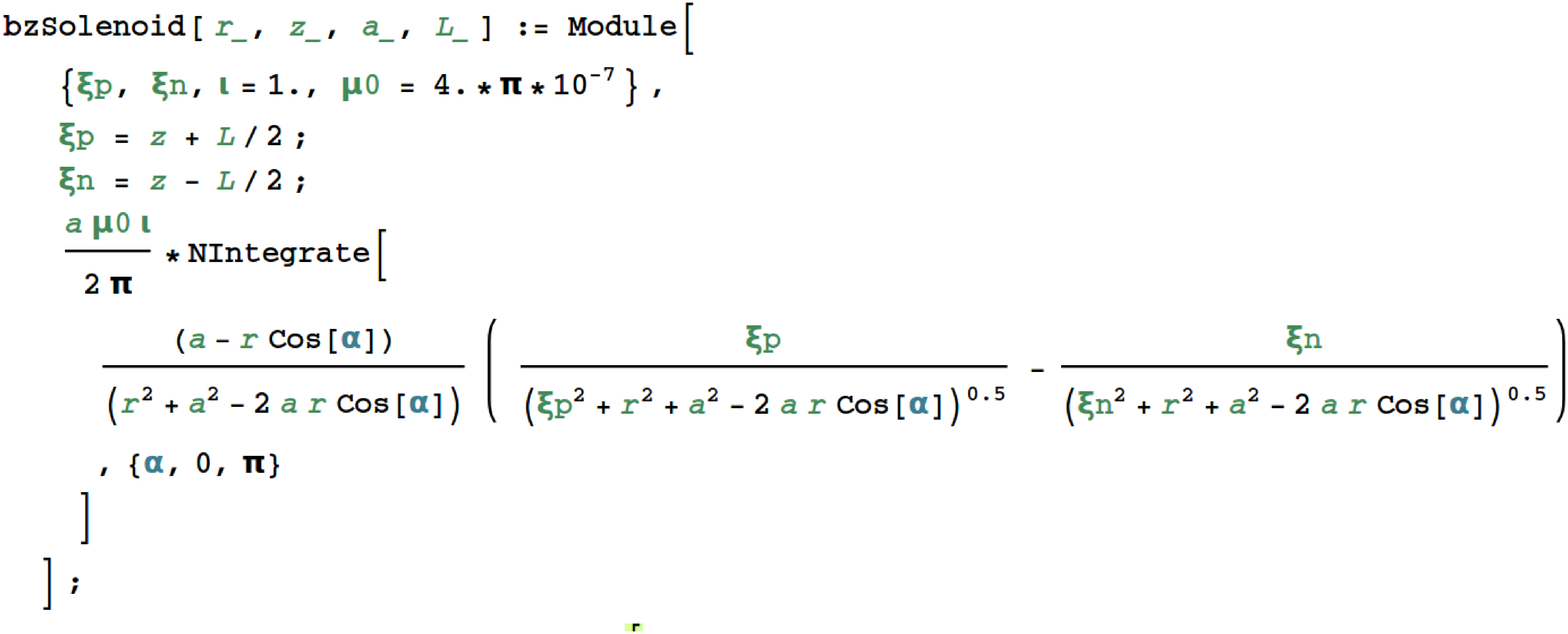}
  \caption{Mathematica code that uses the Mathematica adaptive
    numerical integration function NIntegrate to approximate the magnetic field
    component~$B_z(r,z,a,L)$ given by
    Eq.~(\ref{eq-bz-nasa-1d-integral}) at a point~$(r,z)$ in space,
    for a solenoid of radius~$a$, length~$L$, and current
    density~$\iota=1$.}
  \label{fig-mma-NIntegrate-code-for-bz}
\end{figure}

We used the Mathematica function \verb|NIntegrate| to approximate the
key one-dimensional integrals, Eqs.~(\ref{eq-br-nasa-1d-integral}) and~(\ref{eq-bz-nasa-1d-integral}),
that give the cylindrical components~$B_r$ and~$B_z$ of the magnetic
field generated by a finite solenoid. Details about this function are
available through the freely accessible URL
\verb|reference.wolfram.com/language/ref/NIntegrate.html|, so we
mention briefly that this function takes as its arguments some
integrand in symbolic form and the bounds of the definite integral
(see Fig.~\ref{fig-mma-NIntegrate-code-for-bz}) plus many possible
optional parameters. (We used the default parameters and obtained
acceptable accuracy as mentioned further below.) This function then
uses an unspecified adaptive integration algorithm to
numerically approximate the definite integral to some specified accuracy.

We tested the accuracy of the numerical values returned by
\verb|NIntegrate| for Eqs.~(\ref{eq-br-nasa-1d-integral})
and~(\ref{eq-bz-nasa-1d-integral}) by comparing their values with
analytically equivalent expressions that use entirely different
numerical algorithms for their approximation. Researchers have
shown\cite{Callaghan1960,SolenoidWikiArticle}, that
Eqs.~(\ref{eq-br-nasa-1d-integral}) and~(\ref{eq-bz-nasa-1d-integral})
can be expressed in terms of elliptic integrals\cite{Morse-I-53} as
follows
\begin{eqnarray}
  B_r &=& \frac{\mu_0 \iota}{\pi}
  \sqrt{\frac{a}{r}}
  \left[
    \frac{E(k^2)}{k} + \frac{k^2 - 2}{ 2 k } K(k^2)
  \right]_{\xi_-}^{\xi_+}
 v ,   \label{eq-br-elliptic-integral-expression-nasa} \\
  B_z &=& \frac{\mu_0 \iota}{4 \pi}
  \frac{1}{\sqrt{a r}}
  \left[
    \xi k 
    \left(
      K(k^2)  +
      \frac{ a - r }{ a + r }
      \Pi\left(h^2, k^2 \right)
    \right)
   \right]_{\xi_-}^{\xi_+} ,   \label{eq-bz-elliptic-integral-expression-wiki}
\end{eqnarray}
where the symbols and functions have the following definitions:
\begin{eqnarray}
  k^2  &=& \frac{ 4 a r }{ \xi^2 + (a + r)^2 } , \label{eq-k2-defn} \\
  h^2  &=& \frac{ 4 a r }{ (a + r)^2 + \xi^2 } ,   \label{eq-h2-defn} \\
  K(m) &=& \int_0^{\pi/2} \! \frac{ d\theta }{ \sqrt{1 - m \sin^2\theta} } 
            ,   \label{eq-K-elliptic-integral-defn} \\
  E(m) &=& \int_0^{\pi/2} \! \sqrt{ 1 - m \sin^2\theta } \, d\theta 
            ,  \label{eq-E-elliptic-integral-defn} \\
  \Pi(n,m) &=& \int_0^{\pi/2}
  \frac{ d\theta }{
    \left(1 - n \sin^2 \theta \right)
    \sqrt{1 - m \sin^2 \theta}
    } .   \label{eq-Pi-elliptic-integral-defn}
\end{eqnarray}
The algorithms used by Mathematica to approximate elliptic integrals
numerically use a completely different method than the adaptive
integration algorithm of the \verb|NIntegrate| function. Our
comparison of Eqs.~(\ref{eq-br-nasa-1d-integral})
and~(\ref{eq-bz-nasa-1d-integral}) with
Eqs.~(\ref{eq-br-elliptic-integral-expression-nasa})
and~(\ref{eq-bz-elliptic-integral-expression-wiki}) showed that both
expressions gave relative errors of better than $10^{-8}$ (eight
significant digits) for points near and far from the solenoid's
surface. This accuracy was sufficient for the goals of this paper.

Some further tests we made to validate our results were the following:
\begin{enumerate}
\item We verified that Eq.~(\ref{eq-bz-nasa-1d-integral}) agrees
  accurately with the analytical expression for the magnetic field on
  the axis of a continuous solenoid with end surfaces at~$z=\pm L/2$:
  \begin{equation}
    \label{eq-Bz-on-axis-of-solenoid}
    B_z(r=0,z) = \mu_0 \iota \times \frac{1}{2}
    \left( 
        \frac{ z + L/2 }{ \sqrt{ (z + L/2)^2 + a^2 } }
     -  \frac{ z - L/2 }{ \sqrt{ (z - L/2)^2 + a^2 } }
    \right) .
  \end{equation}
  This expression includes the limiting case $L \to 0$ when the
  solenoid reduces to a current loop of radius~$a$ and again the
  numerical approximations to the continuous solenoid were found to be
  accurate in that limit.

  For~$z=0$, Eq.~(\ref{eq-Bz-on-axis-of-solenoid}) predicts that $B_z
  \approx \mu_0 \iota \left(1 - 2(a/L)^2 + \ldots \right)$ to lowest
  order in~$(a/L)^2$, i.e., the relative error $(B_z(0,0)-\mu_0
  \iota)/(\mu_0 \iota)$ decays as $1/L^2$ for large~$L$. This $1/L^2$
  convergence is similar to what is found for the solenoid's external
  field.

\item We verified that Eqs.~(\ref{eq-br-nasa-1d-integral})
  and~(\ref{eq-bz-nasa-1d-integral})) converge to the corresponding
  components
  \begin{equation}
    \label{eq-magnetic-dipole-B-components}
    B_r(r,z) = \frac{ \mu_0 m }{ 4 \pi } 
     \frac{ 3 r z }{ \left( r^2 + z^2 \right)^{5/2} } , \qquad
    B_z(r,z) = \frac{ \mu_0 m }{ 4 \pi } 
     \frac{ 2z^2 - r^2 }{ \left( r^2 + z^2 \right)^{5/2} }
  \end{equation}
  of the magnetic field of the continuous solenoid's equivalent point
  magnetic dipole at its center~$r=z=0$, with magnetic moment
  \begin{equation}
    \label{eq-equivalent-magnet-moment}
    m = \left( \pi a^2 \right) \iota L ,
  \end{equation}
  when the distance~$d=\sqrt{r^2 + z^2}$ of the point~$(r,z)$ of
  evaluation was large, $d \gg a, L$, compared to the solenoid's
  radius or length. That is, the solenoid's external magnetic field
  correctly converges to that of a point magnetic dipole far from the
  solenoid, with a $1/d^3$ decay in
  magnitude. Eq.~(\ref{eq-equivalent-magnet-moment}) was deduced by
  obtained by comparing Eq.~(\ref{eq-Griffiths-far-field-of-solenoid})
  with the magnetic field of a point dipole $\mu_0 m / (4 \pi r^3)$
  for~$z=0$ in the far-field regime~$r \gg L, a$.

\item We verified that the total magnetic field~${\bf B}$ at points of
  certain symmetry had zero components, even though the numbers added
  to obtain the component were nonzero.

\end{enumerate}

\begin{figure}[ht]
 {\includegraphics[width=4in]{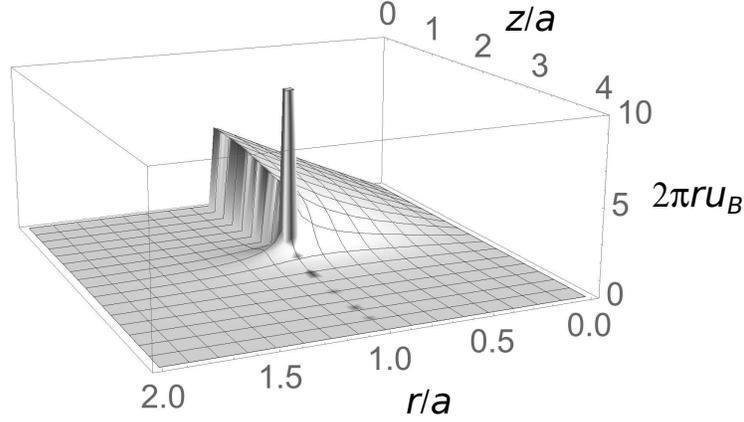}}
\caption{Surface plot of the integrand $(2\pi r) [B^2/(2\mu_0)]$ of
  the magnetic energy, from Eq.~(\ref{eq-magnetic-energy}) for the
  parameters~$a=1$, $L=4$, and~$\iota=1$. The peak at~$r/a=1$ and~$z=
  L/2$ comes from an integrable divergence of the radial
  component~$B_r(r,z)$ of magnetic field (see
  Fig.~\ref{fig-B-field-discontinuities}(a)), while the line of
  discontinuity from $z=0$ to~$z=L/2$ along the line~$r=a$ comes from
  the $z$-component~$B_z$, which changes sign discontinuously as one
  passes from the inside to the outside of the solenoid (see
  Fig.~\ref{fig-B-field-discontinuities}(b)). The two discontinuities
  in the energy integrand lead to a challenging integral to
  approximate numerically.}
\label{fig-surface-plot-mag-energy-integrand}
\end{figure}

We also verified the convergence and accuracy of the 2D integral of magnetic energy, which presents two types of  discontinuities, as shown in Fig.~\ref{fig-surface-plot-mag-energy-integrand}. Firstly, the peak at~$r/a=1$ and~$z=L/2$ comes from a divergence of the radial component~$B_{r}(r,z)$ of the magnetic field. Near the point of divergence,~$B_{r}(r,\pm L/2) \approx{ln|r-a|}$, which does lead to a convergent value for the external magnetic energy. Since the divergence due to the zero-thickness of the continuous solenoid leads to a finite magnetic energy, the magnetic energy of a physical solenoid wound with wires of finite thickness will show results compatible to those presented here. 

%Discussion of how we verified the 2d integral of magnetic energy
%density to be convergent and accurate, the fact that $B_z$ changes
%discontinuously as $r$ increases through $r=a$ for fixed~$z$, and
%$B_r$ diverges to infinity as~$r\to a$ for $z=\pm L/2$ with different
%diverging behaviors on both sides of $r=a$. Show some figures, argue
%that the magnetic energy of a physical solenoid wound from wires will
%show the same result, i.e., our results for an ideal solenoid of a
%pure surface of zero thickness will be consistent with that of a real
%solenoid of a finite thickness and possibly helical windings.

\begin{figure}[ht!]
  \includegraphics[width=4in]{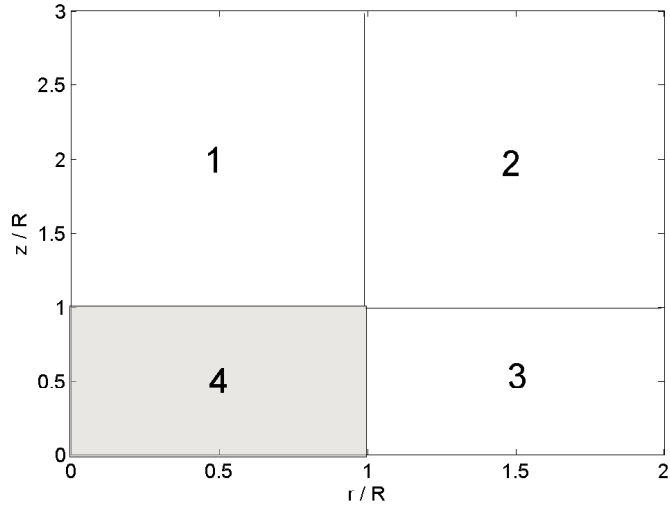}
  \caption{Subdivided region of integration for calculation of
    external and internal energy. We use the symmetry of the solenoid
    to calculate the magnetic field only for the right side of the
    solenoid, $z \ge 0$.  }
\label{fig-external-energy-calculation}
\end{figure}

\section{Adding up radial components~$B_r$ from different solenoids}
\label{app-adding-radial-components}

\begin{figure}[ht!]
\centering 
\includegraphics[width=4in]{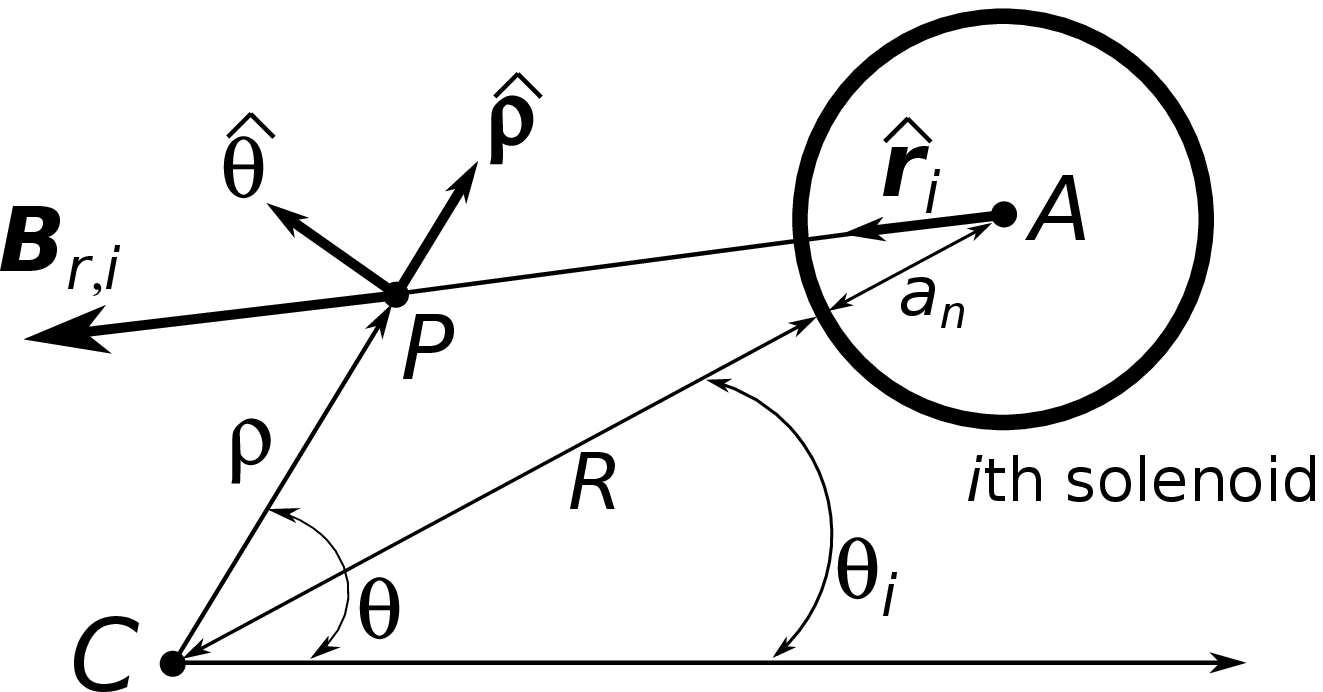}
\caption{Illustration showing how the total magnetic field~${\bf
    B}_{\rm tot}$ at a point~$P=(\rho,\theta,z)$ inside the common
  cylindrical surface~$\rho=R$ with axis at point~C is obtained by
  adding the vectors~${\bf B}_{r,i} = B_{r,i}(r_i,z) \hat{\bf r}_i$ of
  the magnetic field created by the $i$th~solenoid whose axis at
  point~A has the coordinates~$(\rho,\theta) =(R+a_n,\theta_i)$. The
  length~$r_i$ is the radial coordinate of the point~$P$ with respect
  to the solenoid's axis at~A.}
\label{fig-ring-magnetic-field-superposition}
\end{figure}

For both the ring geometry and hexagonal array, one has to add magnetic fields from different directions. We show the main issues for the ring geometry. The calculation of the total magnetic field~${\bf B}(\rho,\theta,z)$
at some point $P = (\rho,\theta,z)$ inside the common cylindrical
surface~$\rho=R$ of the ring of $n$~solenoids described in
Section~\ref{sec:ring-of-solenoids} provides a nice example of how to
change information described in the cylindrical coordinate
system~$(r,\phi,z)$ of one system (a given solenoid) to information
described by the cylindrical coordinate system~$(\rho,\theta,z)$
associated with the common cylindrical surface. The key issue is that
the radial unit vector~$\hat{\bf r}_i$ associated with the
$i$th~solenoid varies with~$i$ as shown in
Fig.~\ref{fig-ring-magnetic-field-superposition}, so one has to
accumulate the components of the vector~$B_{r,i} \hat{\bf r}_i$ along
the unit vectors~$\hat{\boldsymbol{\rho}}$
and~$\hat{\boldsymbol{\theta}}$ of the $(\rho,\theta,z)$ coordinate
system. The $z$-component~$B_z$ is easier to compute since the
$z$~axes of the solenoids are all parallel to the $z$~axis of the
cylindrical surface.

If the radius~$a_n$ of each solenoid is given by
Eq.~\ref{eq-ring-solenoid-radii}, the axis of each solenoid lies on
the circle~$\rho = R + a_n$ and has angular coordinate
\begin{equation}
  \label{eq-theta-n-defn}
  \theta_i = \frac{ 2 \pi }{ n }(i - 1) , \qquad
  \qquad i = 1, \ldots, n .
\end{equation}
If we define the two-dimensional vectors
\begin{eqnarray}
  \hat{\boldsymbol{\rho}}   &=& \bigl( \cos(\theta), \sin(\theta) \bigr) ,\\
  \hat{\boldsymbol{\theta}} &=& \bigl( -\sin(\theta), \cos(\theta)  \bigr) , \\
  \boldsymbol{\rho} &=& \rho \, \hat{\boldsymbol{\rho}} , \label{eq-p-vec} \\
  {\bf a}_i &=& 
     \left( R + a_n \right) \, \bigl( 
       \cos\left( \theta_i \right) , 
       \sin\left( \theta_i \right)
     \bigr)  , \label{eq-ai-vec}
\end{eqnarray}
then the vector $\boldsymbol{\rho} - {\bf a}_i$ points from the axis~A
of the $i$th~solenoid to the point~$P$, its Euclidean length~$r_i$
\begin{equation}
  \label{eq-distance-d}
  r_i = \| \boldsymbol{\rho} - {\bf a}_i \| ,
\end{equation}
is the distance of the point~$P$ to the axis of the $i$th~solenoid,
and so the radial unit vector~$\hat{\bf r}_i$ along the radial
coordinate centered on the $i$th~solenoid is given by
\begin{equation}
  \label{eq-r-radial-vector}
  \hat{\bf r}_i = \frac{ \boldsymbol{\rho} - {\bf a}_i }{ r_i } .
\end{equation}
With this notation, the radial and azimuthal components of the total
magnetic field at the point~$P$ are given by
\begin{eqnarray}
  B_\rho &=&  \hat{\boldsymbol{\rho}} \cdot 
   \sum_{i=1}^n B_{r,i} \hat{\bf r}_i  , \\
  B_\theta &=&  \hat{\boldsymbol{\theta}} \cdot
   \sum_{i=1}^n B_{r,i} \hat{\bf r}_i  ,
\end{eqnarray}
where~$B_{r,i} = B_r(r_i,z)$ is the radial component of the magnetic
field Eq.~(\ref{eq-br-nasa-1d-integral}) at a point a distance~$r=r_i$
from the axis of the $i$th~solenoid.

\begin{acknowledgments}
  We thank Stephen Teithsworth for helpful discussions.
\end{acknowledgments}

\bibliographystyle{unsrt} % entries in order of citation

%\bibliography{main}
\bibliography{solenoid,physics}

\end{document}